%% file: main_V3.tex
\documentclass[%
 amsmath,amssymb,
 aps, physrev,
reprint,
superscriptaddress,
floatfix 
]{revtex4-2}
\input{packages_shortcut}

\begin{document}

\title{Beyond the Kagome Layer: Interlayer Origin of the Flat Band in FeSn}

\author{Shimin Zhang}
\email{zhanshim@iu.edu}
\thanks{These authors contributed equally to this work.}
\affiliation{Department of Chemistry, Indiana University, Bloomington, IN 47405-7102, USA}

\author{Bipasa Samanta}
\thanks{These authors contributed equally to this work.}
\affiliation{Department of Chemistry, Indiana University, Bloomington, IN 47405-7102, USA}

\author{Ho Viet Thang}
\affiliation{The University of Danang, University of Science and Technology, Danang 550000, Vietnam}

\affiliation{Department of Chemistry, Indiana University, Bloomington, IN 47405-7102, USA}
\author{Alexandru Bogdan Georgescu}
\email{georgesc@iu.edu}
\affiliation{Department of Chemistry, Indiana University, Bloomington, IN 47405-7102, USA}

\begin{abstract}
Kagome FeSn exhibits an occupied flat band at its terminated surface, whereas bulk FeSn adopts A-type antiferromagnetic order and does not display the same feature. Here, we combine density functional theory with primitive- and doubled-cell analysis and a double-layer tight-binding model to determine how interlayer electronic coupling and magnetic stacking control flat-band formation in FeSn. We identify an occupied Fe-\(d_{z^2}\)-derived flat-band manifold in the ferromagnetic state that is consistent with the experimentally observed surface feature. Brillouin-zone folding reveals that its flat branch originates from the \(k_z=\pi/c\) sector of the primitive cell, demonstrating that it cannot be understood as an isolated kagome-layer state. Instead, the state depends on coupling between neighboring kagome layers and can be seen as a result of antibonding coupling between nearest neighbor kagome layers. A double-layer tight-binding analysis identifies interlayer Fe–Fe hopping as the dominant microscopic coupling responsible for this behavior, while a contrasting unoccupied flat-band-related manifold is governed primarily by intralayer Fe–Sn hybridization. These results establish interlayer coupling and magnetic stacking as key control parameters for kagome flat bands and highlight how coupling between layers can generate and tune extended correlated-electron states in quantum materials.

\end{abstract}
\maketitle
\section{Introduction}
Kagome materials host characteristic electronic structures arising from their network of corner-sharing triangles, most notably Dirac dispersions and nearly dispersionless flat bands \cite{ghimire2020topology, yin2022topological}. The suppression of kinetic energy in kagome flatbands enhances the role of electronic interactions and provides a platform for correlated magnetism and nontrivial topological phases \cite{wang2024topological, guo2024quantum, ye2018massive, khasanov2024tuning}. Although geometric frustration and destructive interference provide the basic mechanism for kagome flat bands, their dispersion in real materials can be substantially modified by orbital character, electron correlations, reduced dimensionality, and hybridization \cite{kang2020dirac, liu2020orbital, meier2020flat, sales2022flat, checkelsky2024flat, cai2024emergence}. Magnetic order provides an additional degree of freedom because kagome systems support ferromagnetic and antiferromagnetic phases as well as more complex spin textures \cite{chen2024competing, teng2023magnetism, zhang2022electronic, wang2021magnetic, yin2019negative}.

FeSn provides a particularly useful platform for examining how these effects modify kagome flat bands. It consists of Fe kagome layers separated by Sn layers and, in the bulk, exhibits A-type antiferromagnetic order, with ferromagnetic alignment within each Fe kagome plane and antiferromagnetic coupling between neighboring planes~\cite{yosida1951note, haggstrom1975studies}. ARPES and related experiments have established both Dirac-like dispersions and flat-band signatures in FeSn \cite{kang2020dirac, han2021evidence, tao2023investigating, li2022spin, multer2023imaging}. In particular, Kang \textit{et al.} observed a nearly nondispersive occupied band at approximately $-0.23$~eV on the kagome-terminated surface. This feature is prominent under linear-vertical polarization but strongly suppressed under linear-horizontal polarization, indicating a pronounced matrix-element and orbital-symmetry dependence \cite{kang2020dirac}.

The surface-sensitive nature of this observation is important because the electronic environment of a terminated kagome layer differs from that of an interior layer in bulk FeSn. In the bulk, each Fe kagome layer participates in the antiferromagnetic stacking sequence and experiences an approximately inversion-symmetric surrounding environment. At the surface, termination breaks this environment and leaves an uncompensated ferromagnetically aligned kagome layer. Surface and slab studies have shown that this symmetry breaking can substantially reconstruct the kagome-derived electronic states and stabilize surface-localized flat-band features \cite{han2021evidence}. These observations raise a central question: how is the experimentally observed surface flat band related to the magnetic-state-dependent flat-band structure of bulk FeSn, and which microscopic couplings control its formation or suppression?

To address this question, we use density functional theory (DFT) to compare the electronic structures of FeSn in nonmagnetic (NM), ferromagnetic (FM), and antiferromagnetic (AFM) configurations. We find an occupied flat-band manifold in the FM state that is consistent with the experimentally observed surface flat band, and a slightly shifted one in the NM state. Our analysis shows that this state is interlayer in origin and is suppressed by AFM stacking.
We further construct a double-layer tight-binding model to identify the microscopic coupling responsible for its formation, revealing interlayer Fe--Fe hopping as the dominant mechanism. As a contrasting case, we also examine an unoccupied flat-band-related manifold that is governed primarily by intralayer Fe--Sn hybridization and is comparatively insensitive to magnetic stacking. 

The remainder of the paper is organized as follows.
Section~\ref{sec:dft} presents the DFT results, identifying the occupied flat-band manifold associated with the experimentally observed surface state and examining its interlayer origin and dependence on magnetic stacking. Section~\ref{sec:TB} introduces the double-layer tight-binding
model and uses band-tracking metrics and systematic parameter scans to identify the microscopic coupling mechanisms underlying the flat-band dispersions. Section~IV summarizes the roles of interlayer electronic coupling and magnetic stacking in controlling flat-band formation in
FeSn. Computational details are provided in Section~\ref{sec:method}.

\section{DFT band structures and interlayer origin of the flat-band state \label{sec:dft}}
\begin{figure}[h]
    \centering
    \centering\includegraphics[width=1\linewidth]{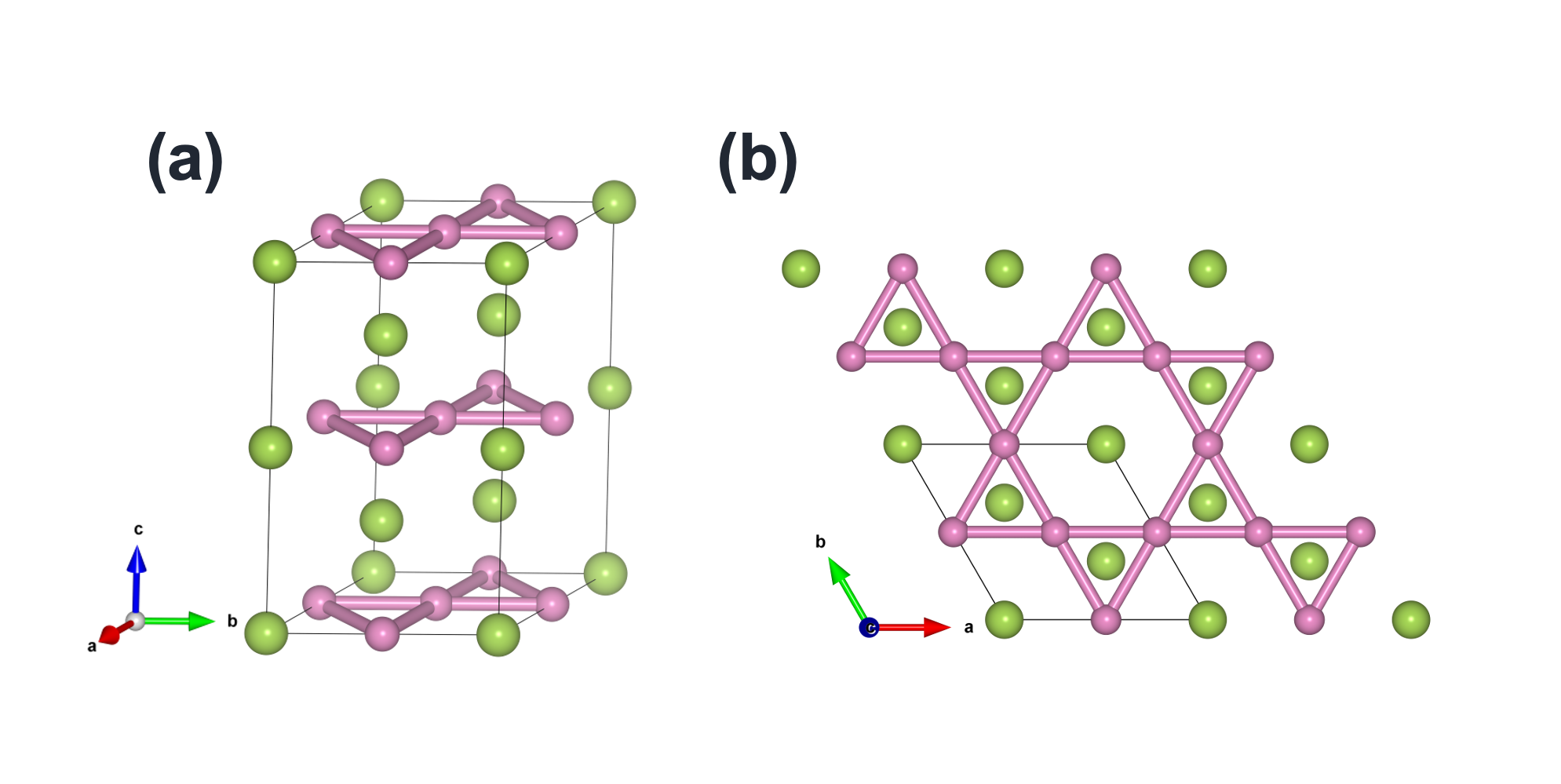} 
    \caption{(a) side view and (b) top view of FeSn kagome lattice with unit cell indicated by black line}
    \label{fig:FeSn_structure}
\end{figure}

The kagome FeSn structure with space group $P6/mmm$ consists of Fe kagome nets filled with Sn atoms and separated by Sn spacer layers~\cite{nial1938rontgenuntersuchung}. To accommodate the layered antiferromagnetic order observed experimentally, we first conduct our DFT calculation with the unit cell doubled along the \textit{c} direction as shown in Fig.~\ref{fig:FeSn_structure}.

\begin{figure*}[ht]
    \centering
    \includegraphics[width=\textwidth]{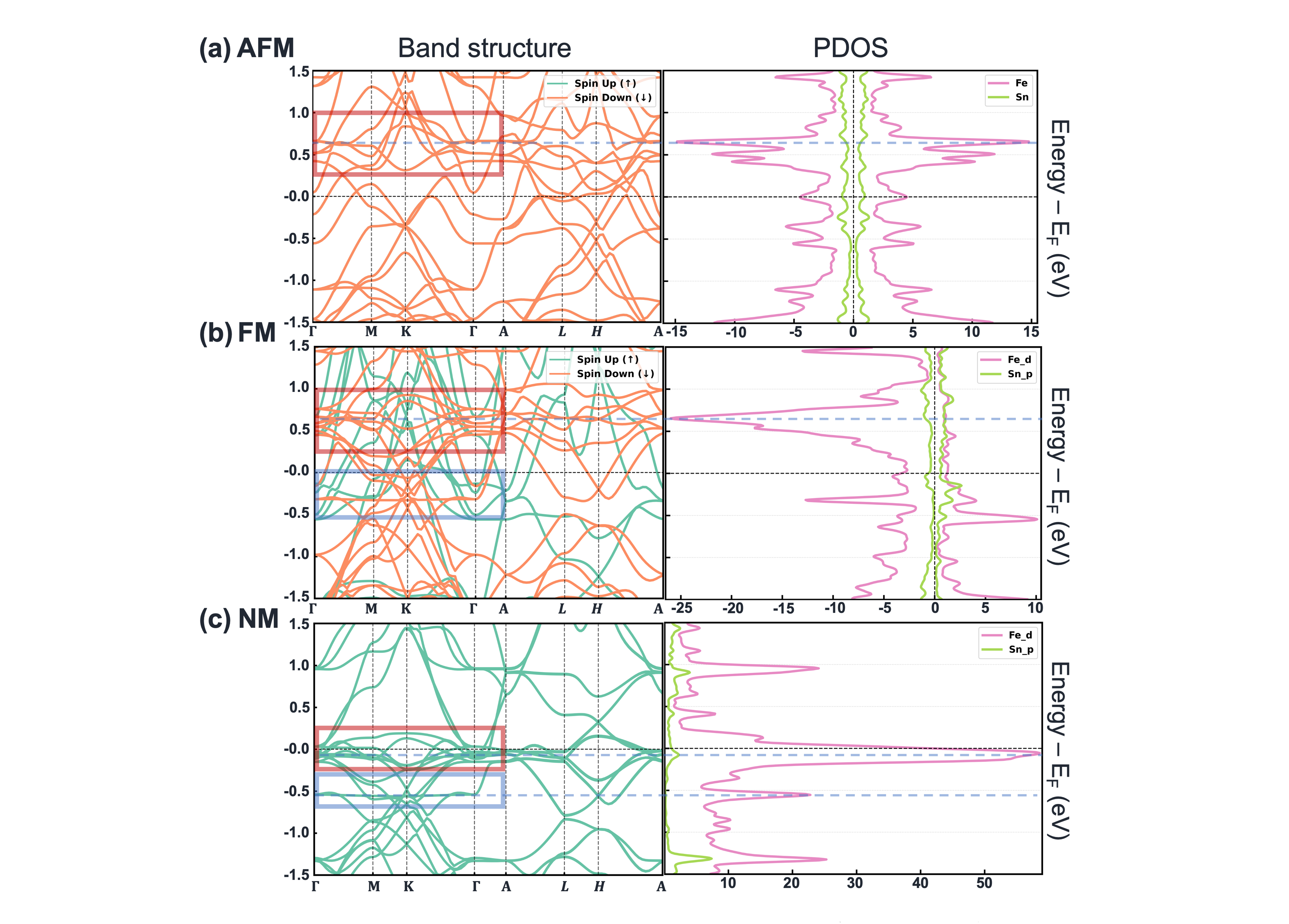} 
    \caption{DFT band structures and projected density of states (PDOS) of kagome FeSn under different magnetic configurations: (a) antiferromagnetic (AFM), (b) ferromagnetic (FM), and (c) nonmagnetic (NM) states. The left panels show the spin-resolved band structures, and the right panels show the corresponding Fe-$d$ and Sn-$p$ projected density of states. The black dashed horizontal line marks the Fermi level. The highlighted regions indicate the flat-band-related features discussed in the text: the A-lift feature is marked by blue rectangles, and the bump-like feature is marked by red rectangles. The blue dashed horizontal lines indicate the energy level of each flat-band feature. }
    \label{fig:DFTband}
\end{figure*}

Among the three magnetic configurations considered, the AFM state is the lowest in energy, consistent with the experimentally established A-type antiferromagnetic order in bulk FeSn \cite{haggstrom1975studies,kang2020dirac}. The FM state lies only $0.041$ eV higher in energy, indicating that the relative alignment of neighboring ferromagnetic kagome layers carries a comparatively small energy cost. In contrast, the NM state is $2.263$ eV above the AFM ground state, demonstrating a strong energetic preference for spin-polarized configurations. The small AFM--FM energy separation further suggests that the interlayer magnetic alignment may be sensitive to changes in the local environment or external perturbations.

Figure~\ref{fig:DFTband} compares the DFT band structures and projected density of states (PDOS) of FeSn in the AFM, FM, and NM configurations. The electronic structure changes substantially with magnetic order. The AFM state exhibits nearly identical spin-resolved bands, consistent with its compensated magnetic structure, whereas the FM state shows pronounced majority- and minority-spin splitting. The NM state is qualitatively different and exhibits a large Fe-$d$ density of states near the Fermi level. Across all three configurations, the low-energy electronic states are dominated by Fe-$d$ orbitals, with smaller contributions from Sn-$p$ states, consistent with previous studies~\cite{masrour2014electronic, xie2021spin}. The large Fe-$d$ density of states at $E_F$ in the NM state further supports its instability toward spin polarization.

More importantly, the FM and NM calculations reveal an occupied, nearly dispersionless manifold that closely resembles the surface flat band observed experimentally (marked in blue box in Figure~\ref{fig:DFTband}). The band extends along the in-plane $\Gamma$–M–K–$\Gamma$ path and shifts in energy toward the out-of-plane A point. In the FM configuration, it occurs exclusively in the minority-spin channel at approximately $-0.32$~eV and is predominantly derived from Fe $d_{z^2}$ states. These characteristics are consistent with the experimentally observed occupied flat near $-0.23$~eV, which has likewise been associated with Fe $d_{z^2}$ character~\cite{kang2020dirac, han2021evidence}. For convenience in the following band-tracking and model analysis, we refer to this characteristic dispersion as the ``A-lift'' feature.

\begin{figure*}[ht]
    \centering
    \includegraphics[width=\textwidth]{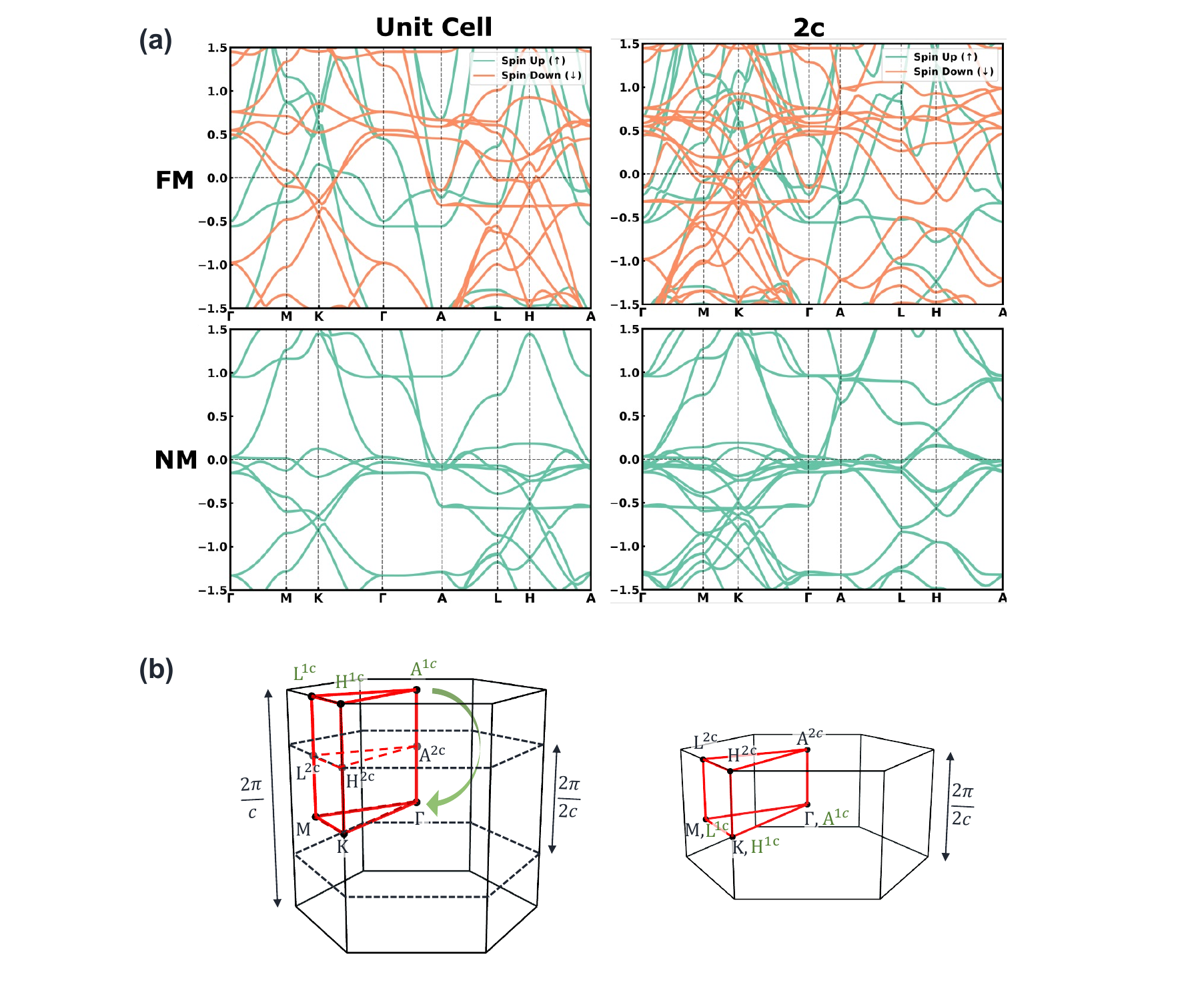} 
    \caption{
    Comparison of the electronic structures and Brillouin zones of the
    primitive $1c$ cell and doubled $2c$ supercell.
    (a) FM and NM band structures calculated using the $1c$ and $2c$ cells.
    (b) Schematic Brillouin zones illustrating the folding along $k_z$.
    Doubling the lattice periodicity along $c$ halves the Brillouin zone in
    the out-of-plane direction. The primitive basal
    $\Gamma\!-\!M\!-\!K\!-\!\Gamma$ plane remains at $k_z=0$, whereas the
    primitive zone-boundary points $A^{1c}$, $L^{1c}$, and $H^{1c}$ at
    $k_z=\pi/c$ fold onto $\Gamma$, $M$, and $K$, respectively, in the
    $2c$ Brillouin zone. The points $A^{2c}$, $L^{2c}$, and $H^{2c}$ define
    the zone-boundary plane of the doubled cell at $k_z=\pi/2c$.
    }
    \label{fig:bandfold}
\end{figure*}

To determine whether the A-lift reflects an isolated-layer state or depends on the electronic relation between neighboring layers, Fig.~\ref{fig:bandfold}(a) compares the FM and NM band structures calculated using the primitive $1c$ cell and the doubled $2c$ supercell. Hereafter, we refer to the $\Gamma\!-\!M\!-\!K\!-\!\Gamma$ plane as the basal plane
and the $A\!-\!L\!-\!H\!-\!A$ plane as the boundary plane of the corresponding Brillouin zone (BZ). For the primitive $1c$ cell, the boundary plane, $A^{1c}\!-\!L^{1c}\!-\!H^{1c}\!-\!A^{1c}$, lies at $k_z=\pi/c$.

Doubling the lattice periodicity along $c$ halves the BZ along $k_z$, folding the primitive boundary plane onto the basal plane of the $2c$ BZ, while the primitive basal plane remains on the same in-plane path [Fig.~\ref{fig:bandfold}(b)]. The flat branch associated with the A-lift appears on the boundary plane in the primitive cell calculation but is folded onto the basal plane in the $2c$ calculation. The flat portion of the A-lift therefore originates from the boundary-plane sector of the primitive cell rather than from the primitive basal plane.  At $k_z=\pi/c$, translation by one primitive lattice vector $c$ introduces the Bloch phase $e^{ik_zc}=-1$, corresponding to a $\pi$ phase shift between neighboring primitive cells. This out-of-phase relation makes the state strongly sensitive to coherent interlayer coupling and shows that the A-lift band cannot be understood solely as an isolated-kagome-layer feature. 

Having established the interlayer character of this state, we return to the magnetic-state comparison in Fig.~\ref{fig:DFTband}. The flat-band manifold is present in the FM and NM configurations but is absent in bulk AFM FeSn, where the continuous basal-plane dispersion is lost. This contrast provides a natural framework for relating the surface-sensitive ARPES observation to the distinct magnetic-layer environments of the terminated surface and the antiferromagnetic bulk.

For the FM and NM configurations, translation by $c$ along the z direction remains a symmetry, so the $2c$ calculation represents BZ folding of the primitive electronic structure. In the AFM state, however, the magnetic periodicity is doubled along $c$, and translation by $c$ is no longer a symmetry of the magnetic structure. Consequently, the primitive basal- and boundary-plane sectors are no longer independent symmetry sectors and
can undergo genuine reconstruction in the doubled magnetic cell. The absence of the A-lift in bulk AFM FeSn therefore cannot be attributed to band folding alone, but instead reflects a genuine reconstruction of the interlayer-dependent state under AFM stacking.

To determine whether the strong interlayer and magnetic sensitivity of the A-lift is generic to flat-band-related states in FeSn, we also examine a second manifold above the Fermi level. This band develops a pronounced dispersion in the in-plane $M\!-\!K\!-\!\Gamma$ region while remaining
comparatively flat in the neighboring regions (marked in red box in Figure~\ref{fig:DFTband}). For convenience in the following analysis, we refer to this characteristic dispersion as the ``bump-like'' feature.

Unlike the A-lift, the bump-like feature remains visible in the NM, FM, and AFM configurations with relatively minor changes in its overall form. Orbital-resolved analysis further distinguishes the two manifolds: whereas the A-lift is predominantly derived from Fe $d_{z^2}$ states, the bump-like band has stronger contributions from in-plane Fe-$d$
orbitals. Because this feature lies above $E_F$, it is not accessible to conventional occupied-state ARPES and therefore does not correspond to the experimentally observed flat band below the Fermi level.

The primitive-cell calculation provides an additional contrast. The bump-like feature is already present on the primitive basal plane at $k_z=0$, where the Bloch phase under translation by $c$ is unity, and retains essentially the same dispersion after cell doubling, apart from the expected band duplication. Thus, unlike the A-lift, this manifold shows much weaker dependence on the out-of-plane phase relation and
magnetic stacking and is primarily associated with the basal-plane sector.

The contrasting orbital characters, magnetic sensitivities, and $k_z$ origins of the two manifolds suggest that they are governed by different microscopic coupling mechanisms. We therefore use the double-layer tight-binding model below to determine which hopping channels control the interlayer-sensitive occupied flat band and the comparatively robust unoccupied counterpart.

\section{Tight-binding analysis of flat-band formation\label{sec:TB}}

\subsection{Double-layer tight-binding model}
\begin{figure}[h]
    \centering
    \centering\includegraphics[width=0.9\linewidth]{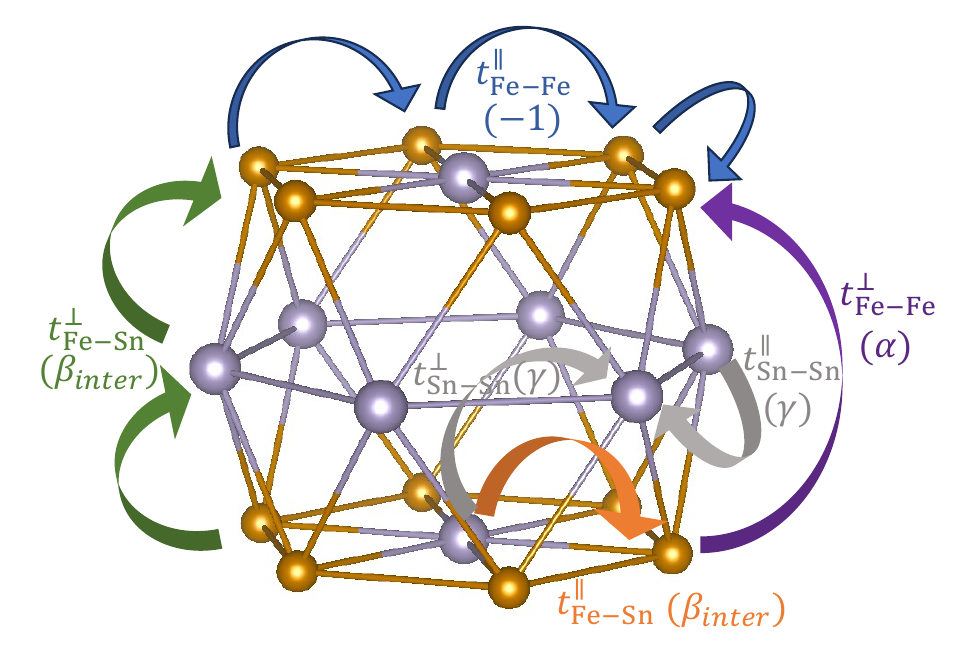} 
    \caption{Double-layer tight-binding structure of FeSn. The colored bonds indicate the hopping channels included in the model: intralayer Fe--Fe hopping (blue), interlayer Fe--Fe hopping (purple), intralayer Fe--Sn hopping (orange), interlayer Fe--Sn hopping (green), and intralayer Sn--Sn hopping (grey). }
    \label{fig:tb_structure}
\end{figure}

To trace the microscopic origin of the flat-band features observed in the DFT band structures, we construct an effective double-layer tight-binding model for kagome FeSn. The model is based on the unit cell doubled along the \textit{c} direction and includes two Fe kagome layers together with the intervening Sn layers, as shown in Fig.~\ref{fig:tb_structure}. This structural setup allows us to separate the relevant hopping and hybridization channels, including intralayer Fe--Sn hybridization, interlayer Fe--Fe hopping, interlayer Fe--Sn hybridization, and Sn--Sn hopping, and to examine how each channel contributes to the formation of the flat-band features.

The model Hamiltonian is written as
\begin{equation}
H = H_{\mathrm{onsite}} + H_{\parallel} + H_{\perp},
\end{equation}
where $H_{\mathrm{onsite}}$ contains the onsite energies, $H_{\parallel}$ includes hopping processes within the same atomic layer, and $H_{\perp}$ includes hopping processes between different layers. Explicitly,

\begin{equation}
H_{\mathrm{onsite}} =
\sum_{l,i} \epsilon_{l,i} c_{l,i}^{\dagger} c_{l,i},
\end{equation}
\begin{equation}
H_{\parallel} =
\sum_{l}\sum_{i,j,\mathbf{R}}
t^{\parallel}_{l i,l j}(\mathbf{R})
c_{l,i,\mathbf{0}}^{\dagger} c_{l,j,\mathbf{R}},
\end{equation}
and
\begin{equation}
H_{\perp} =
\sum_{l\neq l'}\sum_{i,j,\mathbf{R}}
t^{\perp}_{l i,l' j}(\mathbf{R})
c_{l,i,\mathbf{0}}^{\dagger} c_{l',j,\mathbf{R}} .
\end{equation}
Here, $l$ labels the atomic layer, $i$ and $j$ label atomic sites or effective orbitals, and $\mathbf{R}$ is a lattice translation vector. The hopping amplitudes are assigned according to both the atomic species involved and whether the hopping occurs within the same layer or between different layers. We therefore distinguish the hopping channels as
\begin{equation}
\begin{aligned}
&t^{\perp} \in
\left\{
t^{\perp}_{\mathrm{Fe-Sn}},
t^{\perp}_{\mathrm{Fe-Fe}},
t^{\perp}_{\mathrm{Sn-Sn}}
\right\}, \\
&t^{\parallel} \in
\left\{
t^{\parallel}_{\mathrm{Fe-Sn}},
t^{\parallel}_{\mathrm{Fe-Fe}},
t^{\parallel}_{\mathrm{Sn-Sn}}
\right\}.
\end{aligned}
\end{equation}
Here, the superscripts $\parallel$ and $\perp$ denote intralayer and interlayer hopping, respectively, while the subscripts specify the atomic species connected by the hopping process.

The hopping amplitudes also include a distance-dependent decay,
\begin{equation}
t_{ij}(R) =
t_{ij}^{0}
\exp\left[-\frac{ d_{ij}(R)}{\lambda}\right],
\end{equation}
when the distance $d_{ij}(R)$ is within the chosen hopping cutoff, and are set to zero otherwise. This form provides a minimal way to include the geometric effect of the doubled FeSn structure while retaining direct control over the different hybridization channels.

Magnetic order is modeled through layer-dependent Fe onsite energies, which shift the relative alignment of Fe-derived states between adjacent kagome layers. The numerical values of the hopping and onsite parameters used in the scan are described in the following subsection.

\subsubsection{Parameter scan setup}
For the mechanism study, we fix the intralayer Fe--Fe hopping as the reference energy scale, $t^{\parallel}_{\mathrm{Fe-Fe}} = -1.0~\mathrm{eV}$. The remaining hopping channels are then controlled by dimensionless scaling parameters. 
The interlayer Fe--Fe hopping is scaled by $\alpha$,
\begin{equation}
t^{\perp}_{\mathrm{Fe-Fe}} = \alpha t^{\parallel}_{\mathrm{Fe-Fe}},
\end{equation}

the intralayer Fe--Sn hopping is scaled by $\beta_{\mathrm{intra}}$,
\begin{equation}
t^{\parallel}_{\mathrm{Fe-Sn}} = \beta_{\mathrm{intra}} t^{\parallel}_{\mathrm{Fe-Fe}},
\end{equation}
and the interlayer Fe--Sn hopping is scaled by $\beta_{\mathrm{inter}}$,
\begin{equation}
t^{\perp}_{\mathrm{Fe-Sn}} = \beta_{\mathrm{inter}} t^{\parallel}_{\mathrm{Fe-Fe}}.
\end{equation}
The Sn--Sn hopping channels are controlled by a single parameter $\gamma$, which scales both intralayer and interlayer Sn--Sn hopping,
\begin{equation}
t^{\parallel}_{\mathrm{Sn-Sn}} =
t^{\perp}_{\mathrm{Sn-Sn}} =
\gamma t^{\parallel}_{\mathrm{Fe-Fe}}.
\end{equation}

In the parameter scan, the dimensionless hopping parameters are varied over
\begin{equation}
\alpha,\ \beta_{\mathrm{intra}},\ \beta_{\mathrm{inter}},\ \gamma
\in \{-1.0,-0.5,0.0,0.5,1.0\}.
\end{equation}
Magnetic configurations are controlled by the Fe onsite shift $\Delta$. For the NM configuration, $\Delta=0$; for the FM configuration, the two Fe kagome layers are assigned the same onsite shift; and for the AFM configuration, they are assigned opposite shifts, $\Delta$ and $-\Delta$. The magnetic calculations use $\Delta=0.3$ eV. The Sn onsite energy is fixed at
\begin{equation}
E_{\mathrm{Sn}}=-0.8~\mathrm{eV},
\end{equation}
to separate the Fe-derived manifold from Sn-dominated states.

This scan allows us to independently vary the strength and sign of the main hopping channels and to test how interlayer coupling, Fe--Sn hybridization, Sn--Sn hopping, and magnetic layer asymmetry reshape the low-energy band structure.

This parameter scan provides a controlled way to vary the relative strength and sign of the main hybridization channels. By comparing the resulting tight-binding band structures across the parameter grid, we identify which combinations of interlayer coupling, Fe--Sn hybridization, Sn--Sn hopping, and magnetic layer asymmetry reproduce the characteristic flat band features found in the DFT calculations.
\subsubsection{Feature-identification metrics for flat band}
To identify model bands corresponding to the DFT-observed flat-band features, we use a band-continuity tracking and feature-scoring procedure. Since the energy-sorted band index may change near band crossings, continuous bands are first constructed using eigenvector overlaps between neighboring $k$ points. Each tracked band is then evaluated by two scores designed to capture the A-lift and bump-like features, respectively.

All energies are measured relative to the energy of the tracked band at the first $\Gamma$ point,
\begin{equation}
E_{\mathrm{ref}} = E_{\mathrm{band}}(\Gamma_1).
\end{equation}
For the A-lift feature, the score combines the in-plane flatness of the band with the energy lift at the A point:
\begin{equation}
S_{\mathrm{A-lift}} =
2S_{\mathrm{flat}} + \max(0,0.15-L_A),
\end{equation}
where $S_{\mathrm{flat}}$ measures the average deviation of the band from $E_{\mathrm{ref}}$ along the in-plane $\Gamma_1$--M--K--$\Gamma_2$ path, and $L_A=E_A-E_{\mathrm{ref}}$ is the A-point lift. 

For the bump-like feature, the score combines the base-region flatness, the bump height near K, and a roughness penalty:
\begin{equation}
\begin{aligned}
S_{\mathrm{bump}} =
S_{\mathrm{base}}
&+
\max(0,0.1-|H_{\mathrm{bump}}|)\\
&+
0.3\max(0,|H_{\mathrm{bump}}|-1.0)\\
&+
P_{\mathrm{rough}}.
\end{aligned}
\end{equation}

Here, ($S_{\mathrm{base}}$) measures the flatness of the ($\Gamma_1$)--M and ($\Gamma_2$)--A regions, and ($H_{\mathrm{bump}}=E_K-E_{\mathrm{ref}}$) measures the bump height. To suppress sharp point-like features near band crossings, we define
\begin{equation}
P_{\mathrm{rough}} =
\begin{cases}
0, & S_{\mathrm{rough}} < 10^{-3},\\
10, & S_{\mathrm{rough}} \ge 10^{-3},
\end{cases}
\end{equation}
where ($S_{\mathrm{rough}}$) is the mean absolute second finite difference of the tracked band in a local K-centered window. 
\begin{equation}
S_{\mathrm{rough}}=\frac{1}{N_K-2}
\sum_{i\in P_K}
\left|
E_{i+1}-2E_i+E_{i-1}
\right|
\end{equation}
Lower scores indicate better agreement with the target smooth bump-like feature. Bands with poor wave-function continuity or artificial global flatness are penalized separately. The detailed band-tracking algorithm, sampling-point definitions, and penalty terms are provided in the Supplemental Material.

\subsection{Interlayer mechanism of the experimentally relevant flat band}
\begin{figure*}[ht]
    \centering
    \includegraphics[width=1\linewidth]{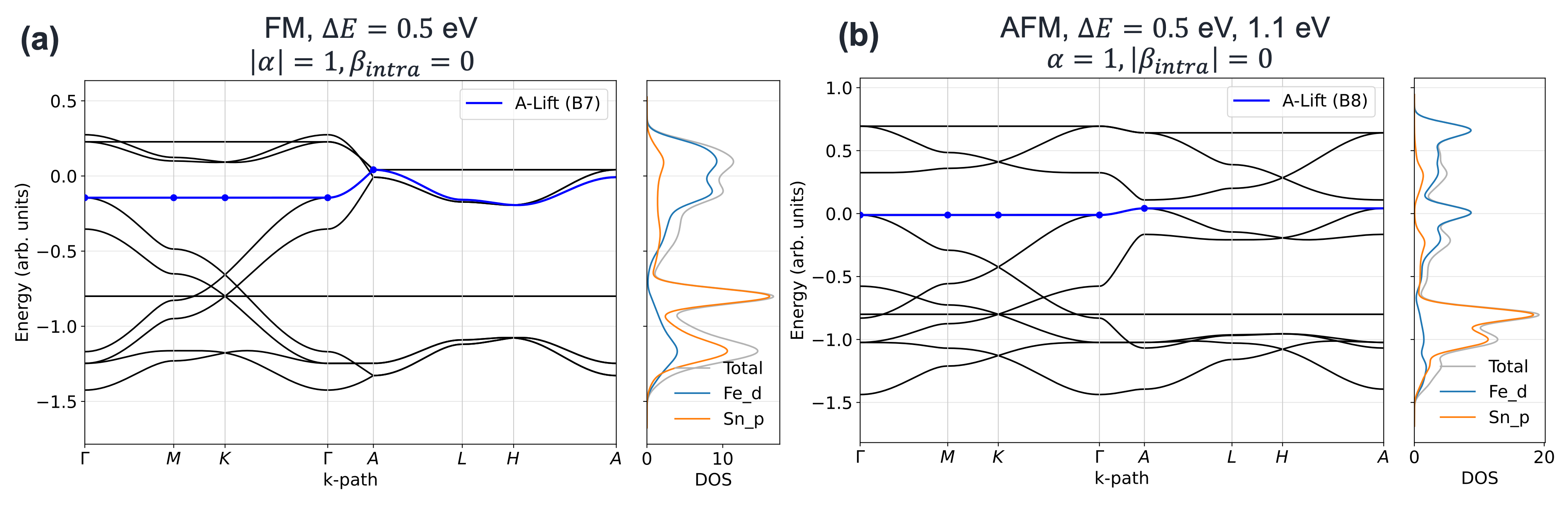}
    \caption{Representative tight-binding candidates for the A-lift flat-band feature. The blue bands indicate the tracked bands selected by the A-lift score, and the projected density of states is shown beside each band structure. (a) FM-like configuration with strong interlayer Fe--Fe hopping, $|\alpha|=1$, and absent intralayer Fe--Sn hopping, $\beta_{\mathrm{intra}}=0$. The selected band remains nearly flat along the basal-plane $\Gamma$--M--K--$\Gamma$ path and shows an energy lift near the A point. (b) AFM-like configuration under the same hopping condition. The layer-dependent Fe onsite energies break the equivalence between the two kagome layers and reduce the A-point lift.}
    \label{fig:alift}
\end{figure*}

The parameter scan identifies strong interlayer Fe--Fe hopping, $|\alpha|=1$, as the common condition across all A-lift candidates. Most candidates occur when intralayer Fe--Sn hybridization is weak or moderate. However, at the smaller Fe--Sn onsite-energy detuning $\Delta E=0.5$~eV, the A-lift can persist even for $|\beta_{\mathrm{intra}}|=1$, provided that strong interlayer Fe--Fe coupling is retained. This shows that weak intralayer Fe--Sn hopping is not a necessary condition for A-lift formation, while its effect depends
on the relative Fe--Sn energy alignment. Strong interlayer Fe--Fe coupling therefore remains the only common microscopic condition across the identified A-lift parameter groups.

Figure~\ref{fig:alift} shows representative A-lift candidates. In the
NM-like and FM-like cases, represented by Fig.~\ref{fig:alift}(a), the
two Fe kagome layers remain equivalent. Strong interlayer Fe--Fe hopping
couples the two layer-derived flat bands and lifts their degeneracy along
the basal plane, while the bands remain degenerate along the boundary
plane. This momentum-dependent layer splitting produces the characteristic
A-lift dispersion. The result is consistent with the strong out-of-plane
phase dependence identified by the DFT folding analysis, indicating that
the A-lift originates from coherent interlayer Fe--Fe coupling whose
effect depends strongly on the out-of-plane Bloch phase.

For the AFM-like configuration in Fig.~\ref{fig:alift}(b), a nearly flat band can still be retained along the basal plane when strong interlayer Fe--Fe hopping is imposed. In the identified AFM-like parameter groups, this occurs for vanishing intralayer Fe--Sn hopping. Introducing opposite layer-dependent onsite shifts breaks the equivalence
of the two kagome layers and reduces the energy lift toward the $A$ point, while largely preserving the basal-plane flat dispersion. Thus, within the model, layer asymmetry primarily suppresses the characteristic A-lift rather than eliminating the underlying flat band.

Importantly, the hopping amplitudes in the tight-binding scan are independent model parameters and are not constrained by the orbital and spin character of a particular magnetic configuration. The AFM-like candidates above therefore establish only that an A-lift can be generated mathematically if sufficiently strong interlayer Fe--Fe hopping against the intralayer Fe-Sn is retained. In real FeSn, however, the effective hopping amplitudes depend on orbital overlap and magnetic alignment. The requirement of strong interlayer Fe--Fe coupling is consistent with the predominantly Fe $d_{z^2}$ character of the A-lift manifold identified by DFT. From this orbital perspective, we hypothesize that the opposite spin alignment of adjacent kagome layers in bulk AFM FeSn reduces the effective coupling between these $d_{z^2}$-derived states, making the strong-$|\alpha|$ regime required for the A-lift inaccessible. This provides a possible physical explanation for why the A-lift is absent in the bulk AFM DFT calculation despite being allowed within the unconstrained parameter space of the model.

\subsection{Contrasting intralayer-controlled flat-band manifold}
\begin{figure*}[ht]
    \centering
    \includegraphics[width=1\linewidth]{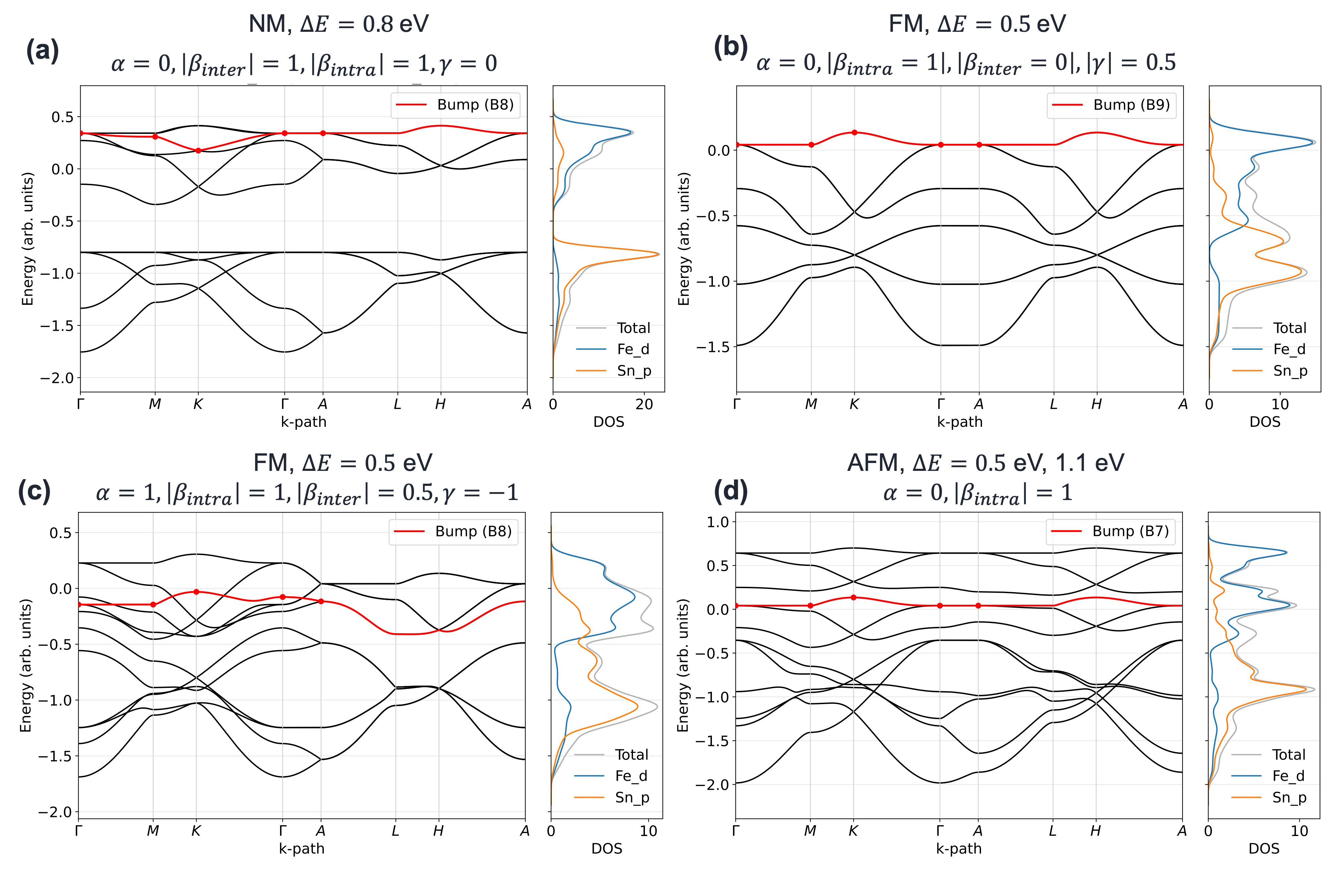}
    \caption{Representative tight-binding candidates for the bump-like flat-band feature. The red bands indicate the tracked bands selected by the bump-score metric, and the projected density of states is shown beside each band structure using gaussian smearing of 0.02 eV. The $\Delta E$ is defined as the on-site energy difference between Fe and Sn. The (a) In the NM-like configuration with ($\Delta E$=0.8) eV, the bump-like feature appears when both intra- and interlayer Fe--Sn hopping are strong, while interlayer Fe--Fe hopping is absent. (b) When the Fe--Sn onsite-energy detuning is reduced to ($\Delta E=0.5$) eV, strong intralayer Fe--Sn hopping alone is sufficient to produce the upper branch of the bump-like feature. (c) Interlayer Fe--Fe hopping and Sn--Sn hopping further tune the dispersion and flatness of the bump-like manifold. (d) In the AFM-like configuration, the bump-like feature persists under layer-dependent Fe onsite energies when the intralayer Fe--Sn hopping remains strong. These representative cases show that the bump-like feature is primarily controlled by intralayer Fe--Sn hybridization, while onsite-energy detuning and interlayer hopping channels modulate its detailed dispersion. }
    \label{fig:bump}
\end{figure*}
In contrast to the interlayer-sensitive A-lift, the bump-like manifold persists across the NM-like, FM-like, and AFM-like configurations. The parameter scan shows that its formation is primarily associated with intralayer Fe--Sn hybridization, consistent with the stronger in-plane Fe-$d$ orbital character identified by DFT. The Fe--Sn onsite-energy
detuning further controls the strength of the resulting dispersion, with smaller $\Delta E$ enhancing Fe--Sn hybridization and producing a more pronounced distortion.

Representative tight-binding results are shown in
Fig.~\ref{fig:bump}. Across the selected parameter groups, strong intralayer Fe--Sn hopping is the common condition associated with the bump-like dispersion, while variations in interlayer Fe--Fe, interlayer Fe--Sn, and Sn--Sn hopping mainly modify its detailed shape and band splitting. This result is consistent with the stronger in-plane Fe-$d$ orbital character identified by DFT and with the presence of the bump-like manifold in the primitive basal-plane sector at $k_z=0$. Together, these results indicate that the bump-like dispersion is governed predominantly by intralayer hybridization rather than by the interlayer coupling that controls the A-lift. The detailed parameter dependence is provided in the Supplemental Material.

This intralayer mechanism also suggests a possible secondary role of surface termination. At the terminated FeSn surface, changes in the local Sn bonding environment may modify the effective intralayer Fe--Sn hybridization. Because stronger Fe--Sn hybridization drives the model toward the bump-like dispersion, a reduction of this coupling at the surface could instead favor a flatter band. Although this effect is not
examined explicitly here, it provides an additional route by which surface termination may modify the flat-band dispersion.

\section{Conclusions}

In conclusion, our results show that the experimentally relevant flat band in FeSn cannot be understood as an isolated kagome layer state. DFT identifies an occupied flat-band manifold in the FM configuration that closely matches the experimentally observed surface state in energy
and orbital character as a result of coupling between two layers. Comparison between the primitive $1c$ and doubled $2c$ cells further traces this flat branch to the boundary plane sector of the primitive Brillouin zone, demonstrating its strong dependence on the phase relation and electronic coupling between neighboring kagome layers. The disappearance of the same manifold in bulk AFM FeSn then shows that magnetic stacking controls whether this interlayer-coupled flat-band condition is realized.

The double-layer tight-binding analysis identifies strong interlayer Fe--Fe hopping as the dominant microscopic coupling underlying this behavior. For equivalent kagome layers, interlayer Fe--Fe coupling produces the momentum-dependent layer splitting associated with the flat-band dispersion, whereas AFM-like layer asymmetry reduces this effect. Together with the predominantly Fe $d_{z^2}$ character obtained
from DFT, these results support a physical picture in which the magnetic alignment of neighboring kagome layers controls the effective interlayer
coupling of the flat-band states. Opposite spin alignment in bulk AFM FeSn may therefore suppress the interlayer coupling required to sustain the experimentally relevant flat-band manifold.

The unoccupied bump-like manifold provides a contrasting case. Unlike the experimentally relevant flat band, it is primarily associated with
intralayer Fe--Sn hybridization and remains comparatively robust across magnetic configurations. This contrast shows that the strong magnetic and interlayer sensitivity of the occupied flat band is not a generic property of all kagome-derived flat-band states. Surface termination may further modify these competing coupling channels through changes in the local magnetic environment and Sn bonding. More broadly, our results establish interlayer electronic coupling and magnetic stacking as central degrees of freedom for controlling flat-band formation in multilayer kagome materials.
Our work highlights the role of interlayer coupling in the formation of extended correlated electron states in quantum materials, as the experimentally observed flatband state can be seen as the result of antibonding coupling between nearest neighbor kagome layers.
\section{Computational setup}\label{sec:method}
Density functional theory (DFT) calculations were performed using the \textsc{Quantum ESPRESSO} open-source software suite~\cite{giannozzi2009quantum, giannozzi2017advanced, giannozzi2020quantum}. The exchange-correlation functional was treated within the generalized gradient approximation (GGA) using the Perdew--Burke--Ernzerhof (PBE) functional~\cite{perdew1996generalized}. We used GBRV ultrasoft pseudopotentials~\cite{garrity2014pseudopotentials}, with valence electrons explicitly treated as Fe ($3p$, $3d$, $4s$) and Sn ($5s$, $5p$). The electronic self-consistency threshold was set to $10^{-8}$, and a plane-wave kinetic-energy cutoff of 40 Ry was used. Gaussian smearing was employed for Brillouin-zone integration.

We use the experimental lattice parameters $a = 5.297~\text{\AA}$ and $c = 8.898~\text{\AA}$~\cite{mikhaylushkin2008high}, with an $11\times11\times6$ $k$-point mesh was used for the doubled $2c$ unit cell. For the additional primitive $1c$-cell calculations used in the Brillouin-zone-folding analysis, an $11\times11\times12$ mesh was used to maintain a comparable sampling density along $k_z$.

\section*{Acknowledgments}
This work was supported by the U.S. Department of Energy, Office of Science, Basic Energy Sciences, under the Early Career Research Program Award DE-SC0026069. The authors acknowledge support from the USAID Partnership for Higher Education Reform project for this work. This work is partially supported by the Lilly Endowment, Inc., through its support for the Indiana University Pervasive Technology Institute Indiana University

\bibliographystyle{apsrev4-2}
\bibliography{Ref_paper}

\end{document}

%% file: packages_shortcut.tex
\usepackage{graphicx} 
\usepackage{url}
\usepackage{hyperref} 
\usepackage[dvipsnames]{xcolor} 
\usepackage{amsmath,amssymb}
\usepackage{svg} 
\usepackage{physics}
\usepackage{comment}
\usepackage{multirow}
\usepackage[normalem]{ulem}